\documentclass[manuscript]{aastex}
\DeclareTextSymbol{\degre}{T1}{6}
\DeclareTextSymbol{\degre}{OT1}{23}
\usepackage{wasysym}

\slugcomment{Submitted to Icarus}

\shorttitle{A primordial origin for the atmospheric methane of Saturn's moon Titan}
\shortauthors{O. Mousis et al.}

\begin{document}


\title{A primordial origin for the atmospheric methane of Saturn's moon Titan\\%[5cm]
}


 \author{
Olivier~Mousis\altaffilmark{1,2},
Jonathan~I.~Lunine\altaffilmark{1},
Matthew~Pasek\altaffilmark{1},
Daniel~Cordier\altaffilmark{3,4},
J.~Hunter Waite, Jr.\altaffilmark{5},
Kathleen~E.~Mandt\altaffilmark{5},
William~S.~Lewis\altaffilmark{5},
and Mai-Julie~Nguyen\altaffilmark{5}}

\altaffiltext{1}{Lunar and Planetary Laboratory, University of Arizona, Tucson, AZ, USA}

\altaffiltext{2}{Universit{\'e} de Franche-Comt{\'e}, Institut UTINAM, CNRS/INSU, UMR 6213, Observatoire des Sciences de l'Univers de Besancon, France}
 
\altaffiltext{3}{Universit{\'e} de Rennes 1, Institut de Physique de Rennes, CNRS, UMR 6251, France}

\altaffiltext{4}{Ecole Nationale Sup{\'e}rieure de Chimie de Rennes, France}

\altaffiltext{5}{Space Science and Engineering Division, Southwest Research Institute, San Antonio, TX 78228, USA}

\begin{abstract}

The origin of Titan's atmospheric methane is a key issue { for understanding the origin of the Saturnian satellite system.} It has been proposed that serpentinization reactions in Titan's interior could lead to the formation of the observed methane. Meanwhile, alternative scenarios suggest that methane was incorporated in Titan's planetesimals before its formation. { Here, we point out that serpentinization reactions in Titan's interior are not able to reproduce the deuterium over hydrogen (D/H) ratio observed at present in methane in its atmosphere, and would require a maximum D/H ratio  in Titan's water ice 30\% lower than the value likely acquired by the satellite during its formation, based on Cassini observations at Enceladus. } Alternatively, production of methane in Titan's interior via radiolytic reactions with water can be envisaged but the associated production rates remain uncertain. On the other hand, a mechanism that easily explains the presence of large amounts of methane trapped in Titan { in a way consistent with its measured atmospheric D/H ratio} is its direct capture in the satellite's planetesimals at the time of their formation in the solar nebula. In this case, the mass of methane trapped in Titan's interior can be up to $\sim$1,300 times the current mass of atmospheric methane.

\end{abstract}


\keywords{Saturn, satellites; satellites, composition; satellites, formation; satellites, atmospheres}



\section{Introduction}

Deuterium is a sensitive indicator of the origin and evolution of planetary atmospheres.  Current deuterium over hydrogen ratios (D/H) measured in the atmospheres of Venus and Mars suggest that these planets were wetter in ancient times than they are at present (L{\'e}cuyer et al., 2000; Krasnopolsky et al., 1998). In the case of Titan, any scenario describing a possible origin for the { observed} methane has to account for its present high D/H atmospheric ratio of  1.32$^{+0.15}_{-0.11} \times 10^{-4}$ (B{\'e}zard et al., 2007), which represents an enrichment of 4.5 to 7.2 times the protosolar value (Cordier et al., 2008). It has been proposed that this observed D/H enhancement could be the result of photochemical enrichment of deuterium through that isotope's preferential retention during methane's photolysis (Pinto et al., 1986; Lunine et al., 1999). However, Cassini-Huygens data have been recently used to reexamine this possibility and it was shown that the photochemical enrichment of deuterium is not sufficient to explain the measured D/H value (Cordier et al., 2008; Mandt et al., 2009). A possible fractionation between CH$_3$D and CH$_4$ during the escape process may slightly enhance the deuterium enrichment by a factor of at most 2.6 times the protosolar value (Cordier et al., 2008), but is not sufficient to explain the observed D/H enhancement over the range of escape values proposed in the literature (Cordier et al., 2008; Mandt et al., 2009). This suggests that D/H in methane is already substantially oversolar when released into the atmosphere of Titan.

Here, we examine the impact of the D/H value in methane on the proposed formation of methane in the interior of Titan from carbon dioxide or carbon grains via the hydrothermal alteration of peridotite (Moody, 1976; Atreya et al., 2006; Oze and Sharma, 2007). In this scenario, the H$_2$ produced during the so-called ``serpentinization'' process in Titan would react with carbon grains or CO$_2$ to produce the methane (Atreya et al., 2006; Oze and Sharma 2007). Our calculations suggest that, { were the atmospheric methane of Titan produced solely by serpentinization reactions in its interior}, the D/H ratio in the primordial water reservoir would not be consistent with the value inferred from the recent Cassini INMS measurement of the D/H ratio in the H$_2$O grains embedded in the Enceladus' plume (Waite et al., 2009). We then discuss the alternative mechanisms that could explain the presence of methane in Titan's atmosphere in agreement with its measured D/H ratio.

\section{Serpentinization reactions in Titan}
\label{serp}

{ In terrestrial oceans, hydrothermal fluids and seawater interact with rock and promote the formation of significant concentration of H$_2$. This process involves peridotite, such as olivine [(Mg, Fe)$_2$SiO$_4$] and pyroxene [(Mg, Fe)SiO$_3$], whose hydratation leads primarily to the formation of molecular hydrogen [H$_{2(\rm aq)}$], together with serpentine [(Mg, Fe)$_3$Si$_2$O$_5$(OH)$_4$], Mg-brucite [Mg(OH)$_2$] and magnetite [Fe$_3$O$_4$].
%via the following reaction (Moody, 1976; Atreya et al., 2006):
%\begin{eqnarray}
%\rm peridotite~(olivine/pyroxene) +  water \rightarrow \rm hydrogen + serpentine  + brucite + magnetite.	\nonumber 
%\end{eqnarray}
Key geochemical reactions are (Moody, 1976; Atreya et al., 2006):
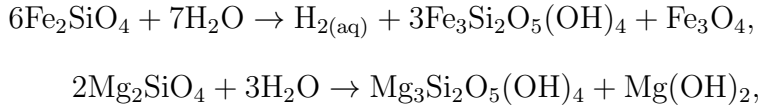
\begin{eqnarray}
\rm 6Fe_2SiO_4 + 7H_2O \rightarrow H_{2(aq)} + 3Fe_3Si_2O_5(OH)_4 + Fe_3O_4,  \nonumber \\
\rm 2Mg_2SiO_4 + 3H_2O \rightarrow Mg_3Si_2O_5(OH)_4 + Mg(OH)_2,  \nonumber
\end{eqnarray}
followed by 
\begin{eqnarray}
\rm 2Fe_3Si_2O_5(OH)_4 + 6Mg(OH)_2 \rightarrow 2H_{2(aq)}+2Mg_3Si_2O_5(OH)_4 + 2Fe_3O_4 + 4 H_2O. \nonumber
\end{eqnarray}}
{ On Titan, the subcrustal liquid water ocean would constitute the free-water reservoir susceptible to interaction with rocks assumed to constitute the core of the body (Tobie et al., 2006).} Serpentinization of Titan's core may have happened during the early history of the satellite when all the ice was molten, or may have happened over geological history as liquid water circulates through the outer parts of the warm rocky core. Because the core of the cooling Titan was cut off early on from the ocean by a growing layer of high-pressure ice whose thickness at present is several hundred kilometers (Tobie et al., 2006), H$_2$ may more likely have been produced at early epochs of the satellite's evolution. CH$_4$ would then be formed from the reaction of H$_2$ with CO or CO$_2$ via Fischer-Tropsch-type synthesis (Anderson, 1984; Matson et al., 2007) or from the heating of graphite- or carbonate-bearing rocks in presence of H$_2$ (Giardini and Salotti, 1968; McCollom, 2003). The methane produced would then saturate the deep ocean (Tobie et al., 2006; Fortes et al., 2007), and would be transported towards the surface of Titan via ammonia-water pockets which erupt through the ice shell and lead to cryovolcanism (Mitri et al., 2006). Alternatively, the methane coming from the saturated ocean could diffuse through the solid icy shell to the surface or could be stored in clathrates formed at the ocean-ice shell interface that would be destabilized and melt by ascent of hot thermal plumes (Tobie et al., 2006).

\section{D/H ratio in the produced methane}
During serpentinization reactions, residual water is deuterium-enriched at the expense of the initial reservoir of free-water. The fractionation factor, $\alpha_{r-w}$, between OH-bearing minerals and water is then (L{\'e}cuyer et al., 2000):

\begin{equation}
\label{frac}
\alpha_{r-w} = \frac {R^f_r}{R^f_w},
\end{equation}

{ \noindent where $R^f_r$ and $R^f_w$ are the D/H ratios in the OH-bearing minerals and the residual water, respectively.}

The effect on the D/H ratio of the residual water of a D/H fractionation between the initial water and hydrated peridotite can be readily tested with the following mass balance equation that describes a batch equilibrium mechanism of both hydration and isotopic fractionation between given masses of rock and water:

\begin{equation}
\label{mass_bal}
M^i_w X^i_w R^i_w + M^i_r X^i_r R^i_r = M^f_w X^f_w R^f_w + M^f_r X^f_r R^f_r,
\end{equation}

\noindent where $M$ is the mass, $X$ the mass fractions of hydrogen in water ($X^i_w$ = $X^f_w$ = 1/9) or rock ($X^f_r$ = 4/277), $R$ the D/H ratios before hydration reactions ($i$) and at batch equilibrium ($f$). If we insert Eq. 1 into Eq. 2 and neglect the likely small amount of water in primordial peridotite of Titan ($X^i_r$ = 0), Eq. 2 becomes:

\begin{equation}
\label{mass_bal2}
R^i_w = \frac{(\frac{M^f_w X^f_w}{\alpha_{r-w}} + M^f_r X^f_r) R^f_r}{M^i_w X^i_w}.
\end{equation}

\noindent Here, we postulate that the D/H ratio in the methane initially released from the interior is that acquired by hydrated rocks once equilibrium is reached during serpentinization reactions. The D/H ratio acquired by hydrated rocks would then be preserved in the hydrogen produced from the alteration of peridotite and used in the recombination of CH$_4$.

The initial mass $M^i_w$ of the free-water reservoir is expected to range between $1.79 \times 10^{22}$  and $6.73 \times 10^{22}$ kg. These values correspond respectively to the current and initial masses of the liquid water ocean (Grasset and Pargamin, 2005), whose volume has decreased with time during the cooling of Titan (Tobie et al., 2006). Assuming that methane has been present continuously in the atmosphere of Titan since 4.5 Gyr and destroyed over Titan's history at the current photolytic destruction rate of $6.9 \times 10^{13}$ molecules m$^{-2}$ s$^{-1}$ (Vuitton et al., 2008), the total mass of expelled methane is $\sim$ $2.17 \times 10^{19}$ kg. Assuming that the hydrogen from the methane is derived from serpentinization reactions only, the equivalent mass of produced H$_2$ is then $\sim$ $5.42 \times 10^{18}$ kg. Depending on the composition of peridotite, key geochemical reactions (Moody, 1976; Charlou et al., 2002) suggest that 4.5 moles of serpentine [Mg$_3$Si$_2$O$_5$(OH)$_4$] are produced per mole of H$_2$ released during serpentinization reactions. Hence, the maximum mass of serpentine $M^f_r$ is then $3.38 \times 10^{21}$ kg, namely about 5\% of Titan's rock mass, over 4.5 Gyr, including 10 wt\% of H$_2$O extracted from the free-water reservoir (L{\'e}cuyer et al., 2000). Because $M^i_w \simeq M^f_w \gg M^f_r$  and $X^i_w \gg X^f_r$, Eq. \ref{mass_bal2} can be simplified as follows:

\begin{equation}
\label{water}
R^i_w \simeq \frac{R^f_r}{\alpha_{r-w}}.
\end{equation}

Equation 4 is of the form of Eq. 1 but concerns R$^i_w$ rather than R$^f_w$. It remains valid independent of the production timescale of methane in the interior of Titan because the mass of hydrated rocks is still small compared to that of the free-water. { Note that assuming an episodic methane outgassing would have led us to a lower estimate of the maximum mass of formed serpentine, hence leading to no change in this condition.} We consider two extreme values, 0.95 and 1.03, of the hydrogen fractionation $\alpha_{r-w}$ between serpentine and the free-water reservoir in the literature based on laboratory and field data made at temperatures ranging between 298 and 773 K (Wenner and Taylor, 1973; Sakai and Tsutumi, 1978; Vennemann and O'Neil, 1996). Finally, we consider two different cases for the value of $R^f_r$. In the first case, we neglect the effect of deuterium photochemical enrichment in the atmosphere of Titan and we consider that $R^f_r$ is equal to $1.32 \times 10^{-4}$, namely the nominal D/H ratio measured by Cassini in Titan's atmosphere (B{\'e}zard et al., 2007). In the second case, we consider the influence of photochemistry that might have enriched the primordial D/H ratio in methane by a maximum factor of 2.6 (Cordier et al., 2008) and then set $R^f_r$ equal to $5.08 \times 10^{-5}$.

Table 1 shows the range of values of $R^i_w$ needed in the primordial water reservoir to explain the D/H ratio observed in the atmospheric methane of Titan. When neglecting the possibility of photochemical enrichment of deuterium in Titan's methane, the values of  $R^i_w$ are about 10--20\% lower than that corresponding to Standard Mean Ocean Water (V-SMOW) (D/H = $1.56 \times 10^{-4}$ (Craig, 1961)). The discrepancy becomes even higher when using values of $R^f_r$ that take into account the possibility of deuterium enrichment in Titan's methane. In this case, the required values of $R^i_w$ are more than 60\% lower than the V-SMOW value. Moreover, the values of $R^i_w$ calculated with Eq. 4 are upper estimates because we make the assumption that the mass of the free-water reservoir in contact with rocks, i.e. that of the liquid water ocean, remains constant during serpentinization in Titan. However, the progressive cooling of the satellite implies the shrinking of the water reservoir available for isotopic exchange with rocks, and consequently significant reduction of $M^f_w$  compared to $M^i_w$. Assuming that the mass of the free-water reservoir has decreased from $6.73 \times 10^{22}$  to $1.79 \times 10^{22}$ kg during the serpentinization process and using Eq. \ref{mass_bal2}, the values of $R^i_w$ range now between $1.34 \times 10^{-5}$ and $3.80 \times 10^{-5}$. Therefore, in all cases evaluated, the initial D/H in the primordial water of Titan's interior is lower than V-SMOW.

\section{Discussion}
\label{discussion}

It is unlikely that the D/H ratio in the primordial water ice accreted by Titan could be as low as the V-SMOW value. Indeed, the INMS instrument aboard the Cassini spacecraft has recently measured D/H in water in the icy grains embedded in the vapor plumes of Enceladus (Waite et al., 2009). It shows that the D/H ratio in the water ice expelled from Enceladus is about $2.9^{+1.5}_{-0.7} \times 10^{-4}$, a value close to that measured in comets (Bockel{\'e}e-Morvan et al., 2004). This implies that the planetesimals accreted by Enceladus were initially produced in the outer solar nebula prior to having being embedded in Saturn's subnebula. Indeed, the D/H ratio of icy planetesimals condensed in an initially dense and warm Saturn's subnebula should be lower than the measured value because an isotopic exchange would have occurred between H$_2$ and H$_2$O in the gas phase of the subdisk, thus decreasing the degree of deuteration acquired by water in the solar nebula (Horner et al., 2008). Since the temperature and pressure conditions are expected to have decreased with increasing distance from Saturn when the latter was surrounded by satellite-forming material (Alibert and Mousis, 2007; Barr and Canup, 2008), { then the planetesimals accreted by Titan were also formed in the outer solar nebula (not in Saturn's subnebula) and should have the same D/H ratio in water as is in the building blocks of Enceladus.} This value is then in conflict with the one required by the hypothesis that serpentinization reactions occurred in Titan's interior. { Even when error bars are taken into account, the maximum D/H ratio (1.55 $\times$ 10$^{-4}$) possessed by  primordial water ice --if CH$_4$ comes entirely from serpentinization reactions --should be at least 30\% lower than the minimum value (2.2 $\times$ 10$^{-4}$) of the D/H ratio measured at Enceladus by the INMS instrument aboard Cassini.}

A word of caution must be given about our estimate of the D/H ratio in methane produced via serpentinization reactions. Our calculations are based on the assumption that no D-fractionation occurred between the produced H$_2$ and the formed serpentine, and during the production of CH$_4$ from H$_2$ and other C-bearing compounds.  We have also assumed that no D-fractionation has occurred during the transport of CH$_4$ towards the surface of the satellite. Unfortunately, except recent laboratory work that shows an extremely light D-fractionation when CH$_4$ is trapped in clathrates ({ the difference in $\delta$(D)\footnote{$\delta$(D) = ($\frac{\rm(D/H)_{sample}}{\rm(D/H)_{standard}}$- 1) 
$\times$ 1000, where D is deuterium, and the standard is V-SMOW (Craig, 1961).} is lower than 10$\permil$ (Hachikubo et al., 2008)}), all the other points still remain unconstrained. 

Interestingly enough, substantial quantities of abiogenically produced CH$_4$ have been observed on Earth within Precambrian rocks of the Canadian Shield, the Fennoscandian Shield and the Witwatersrand basin in South Africa (Sherwood Lollar et al., 1993; Sherwood Lollar et al., 2008). These samples have been found to be deuterium impoverished compared to the V-SMOW value ($\delta$(D)$_{\rm CH_4}$ up to -450$\permil$). However, the observed amplitude of D-fractionation must be tempered by the fact that ground waters are already D-depleted, even if the magnitude of this depletion is lower\footnote{An inventory of the $\delta$(D) values in water in the main terrestrial reservoirs is given in Table 1 of L{\'e}cuyer et al. (1998). $\delta$(D) can be up to -80$\permil$ in the mantle or up to -400$\permil$ in the ice sheets.} than that observed for CH$_4$. In fact, the exact mechanism that led to the formation of these D-depleted CH$_4$ collected samples on Earth is not yet firmly established. Serpentinization has been invoked but, as mentioned above, there is no laboratory experiment that shows the fractionation effects induced by this process on the produced H$_2$. An alternative mechanism, which works experimentally and can explain the most D-light CH$_4$ samples collected on Earth, is the production of H$_2$ from radiolytic reaction with H$_2$O, due to the decay of radioactive elements (U, Th, K etc) (Lin et al., 2005). The $\delta$(D) of H$_2$ produced by radiolysis in laboratory can be up to -500$\permil$ and is then compatible with the observed D/H ratio in Titan's atmospheric methane. However, it is difficult to quantify the production rate of radiolytic H$_2$ in the interior of Titan and compare it to the mass of H$_2$ needed to form the atmospheric CH$_4$. Indeed, the production rate of radiolytic H$_2$ on Earth depends on the concentration of radiogenic elements measured in the rocks and also on the porosity of the mineral matrix (Lin et al., 2005). In particular, the presence of an open porosity in the matrix favors the circulation of free water and increases the efficiency of radiolysis. In the case of Titan, the porosity of the core must be almost zero due to the high pressure (tens of kilobars) exerted by the surrounding layers.  Also, the concentration of radiogenic elements in the mineral matrix in Titan's core is unknown.

On the other hand, a mechanism that easily explains the presence of large amounts of methane trapped in Titan's interior is its direct capture in the building blocks of the satellite at the time of their formation in the solar nebula. Indeed, the CH$_4$/H$_2$O mass ratio in the planetesimals produced in Saturn's feeding zone has been estimated to be $\sim$ 2.2--3.4 $\times$ 10$^{-3}$ (Mousis et al., 2009a). The mass of methane thus trapped in Titan's interior is $\sim$ 1.5--2.3 $\times$ 10$^{20}$ kg and corresponds to $\sim$ 852--1,307 times the current mass of atmospheric methane determined with the data from the Huygens Atmospheric Structure Instrument (Cordier et al., 2008). Moreover, it has been shown that the D/H ratio in the methane acquired in this way could be consistent with the implications of existing atmospheric measurements (Mousis et al., 2002; Cordier et al., 2008; Mandt et al., 2009). Because methane falling into the solar nebula from the interstellar medium was highly enriched in deuterium when compared with the infalling molecular hydrogen, a reversible gas phase isotopic exchange between these two nebular components led to a slow but steady reduction in the deuterium fraction of the methane (Mousis et al., 2002). Once the temperature of the nebula dropped, methane condensed or was clathrated by the available crystalline water ice in the feeding zone of Saturn, halting the reaction and fixing the D/H value acquired in the gas phase (Mousis et al., 2002). These ices agglomerated in the feeding zone of Saturn and formed the building blocks of Titan in the Saturn's subnebula. As methane was released over time into the atmosphere of Titan, a progressive photochemical enrichment of deuterium may have occurred, the magnitude of which is dependant on the time elapsed since the last major outgassing from the interior as well as the nature (episodic or continuous) of the outgassing (Cordier et al., 2008; Mandt et al., 2009). 

This scenario is robust because it does not preclude processes such as the partial devolatilization of planetesimals during their migration within Saturn's subnebula that would explain Titan's observed CO and noble gas deficiencies (Alibert and Mousis, 2007; Mousis et al., 2009b). Moreover, since Enceladus is expected to be formed from the same building blocks as Titan and should also contain large amounts of methane originating from the solar nebula, the measurement of the D/H ratio in the CH$_4$ found in its plumes could provide the initial D/H ratio acquired by primordial methane at the time of its trapping in the solar nebula, and thus information on the efficiency of the deuterium photochemical enrichment that occurred in Titan's atmosphere.

\acknowledgments
This work was supported in part by the French Centre National d'Etudes Spatiales. Support from the Cassini project is also gratefully acknowledged. We acknowledge Barbara Sherwood Lollar for helpful discussions about her work. Many thanks to Ralf Jaumann and an anonymous Referee whose comments have helped us to improve our manuscript.

%% ------------------------------------------------------------------------ %%
%
%  REFERENCE LIST AND TEXT CITATIONS
%
% Either type in your references using

\clearpage
\begin{table}[h]
\centering \caption{Values of $R^i_w$ calculated as a function of the D/H ratio adopted in the methane outgassing from the interior of Titan and for two extreme values of the hydrogen fractionation factor $\alpha_{r-w}$ between serpentine and water (see text).}
\begin{center}
\begin{tabular}{lcc}
\hline 
\hline
							& $\alpha_{r-w} = 0.95$	& $\alpha_{r-w} = 1.03$		\\
\hline
$R^f_r = 5.08 \times 10^{-5}$  		& $5.36 \times 10^{-5}$	&	$4.92 \times 10^{-5}$	\\
$R^f_r = 1.32 \times 10^{-4}$		& $1.39 \times 10^{-4}$	&	$1.28 \times 10^{-4}$	\\
\hline
\end{tabular}
\end{center}
\label{DH}
\end{table}

\end{document}